# The Potential of Combining Thermal Scanning Probes and Phase-Change Materials for Tunable Metasurfaces

Ann-Katrin U. Michel,* Sebastian Meyer, Nicolas Essing, Nolan Lassaline, Carin R. Lightner, Samuel Bisig, David J. Norris, and Dmitry N. Chigrin*

**ABSTRACT:** Metasurfaces allow for the spatiotemporal variation of amplitude, phase, and polarization of optical wavefronts. Implementation of active tunability of metasurfaces promises compact flat optics capable of reconfigurable wavefront shaping. Phase-change materials (PCMs), such as germanium telluride or germanium antimony telluride, are a prominent material class enabling reconfigurable metasurfaces due to their large refractive index change upon structural transition. However, commonly employed laser-induced switching of PCMs limits the achievable feature sizes and thus, restricts device miniaturization. Here, we propose thermal scanning-probe-induced local switching of germanium telluride to realize near-infrared metasurfaces with feature sizes far below what is achievable with diffraction-limited optical switching. Our design is based on a planar multilayer stack and does not require fabrication of protruding dielectric or metallic resonators as commonly applied in the literature. Instead, we numerically demonstrate that a broad-band tuning of perfect absorption could be realized by the localized and controlled tip-induced crystallization of the PCM layer. The spectral response of the metasurface is explained using simple resonance mode analysis and numerical simulations. To facilitate experimental realization, we provide a detailed theoretical description of the tip-induced crystallization employing multiphysics simulations to demonstrate the great potential for fabricating compact reconfigurable metasurfaces. Our concept allows for tunable perfect absorption and can be applied not only for thermal imaging or sensing, but also for spatial frequency filtering.

## INTRODUCTION

Metasurfaces are flat optical elements constructed from a dense array of strongly scattering sub-wavelength nanostructures. Since each such nanosized metallic or semiconductor scatterer can be designed individually, a collective shaping of the light waves based on the tailored local response of individual scatterers is possible. With a generalization of Snell's law,[01] metasurfaces can not only substitute bulky optical components with ultra-thin equivalents, but may also introduce new functionalities (phase discontinuities, anomalous reflection and refraction, *etc.*).[02,03] However, the shape of the wavefront is defined by the metasurface design, including material selection and geometry, and is fixed after fabrication. Active post-fabrication control requires the tunability of the optical response of each metasurface element.[04,05] One common approach for active metasurfaces is to capitalize on the change of the material polarizability – either of the scatterer or its surroundings. This can be achieved, for example, by modulation of the charge density in doped semiconductors (*e.g.* GaAs) or graphene,[06] by changing the state of liquid crystals adjacent to metallic antennas,[07] or by including phase-transition (*e.g.* VO$_2$)[08,09] or phase-change [*e.g.* (GeTe)$_x$(Sb$_2$Te$_3$)$_{1-x}$][10] materials in the metasurface. Mechanical tuning with stretchable substrates or actuators,[11] chemical reactions at the scatterers,[12] or enhanced optical nonlinearities[13] are other possible approaches for post-fabrication tunability.

Phase-change materials (PCMs) are among the best-suited materials to provide tunability of metasurfaces due to the property contrast between their amorphous (A) and crystalline (C) phase.[14] In contrast to phase-transition materials (*e.g.* VO$_2$), PCMs feature non-volatile states and thus, no energy is needed to maintain the material properties. The structural change is accompanied with a refractive index change $\Delta n = |n_C - n_A|$ of up to 3.6.[10,15] The absorption in PCMs for energies above the bandgap $E_G$ (Ge$_2$Sb$_2$Te$_5$: $E_{G,A}$ = 0.77 eV and $E_{G,C}$ = 0.48 eV) facilitates their usage in the mid-infrared (MIR) region,[10,14] but newly identified PCMs and deliberate device designs allow for applications in the near-infrared (NIR) and even the visible spectral range.[16,17] The phase change from the amorphous to the crystalline phase can be induced by annealing,[18] pressure,[19] or electrical[20] or optical pulses.[21-23] However, reversible switching can only be achieved using the last two. The repeatability of the phase change for more than one billion cycles[14] has been crucial for different data-storage and processing applications, such as random-access memories[24] or brain-inspired computing.[25,26] While the thermal parameters for the phase change need to be chosen carefully, a thorough thermodynamic analysis can allow for controlling and predicting the volumetric structural transition.[23,27]

In photonics, PCMs are either applied in hybrid metasurfaces comprised of plasmonic nanostructures with the PCM in their close proximity [10,15,23] or in all-dielectric metasurfaces with PCM-based scatterers.[28-30] The resonances observed in these two different concepts are based on excitation of plasmonic (conduction current) or Mie-type (displacement current) modes, respectively. All-dielectric metasurfaces are motivated by mitigating substantial Ohmic losses encountered in noble metals. A different approach involves the excitation and tuning of phonon polaritons in silicon carbide (SiC) or hexagonal boron nitride (hBN), which feature very low losses.[22,31,32] While the dielectric function of most PCMs make them an ideal platform for photonic applications, the traditional switching approaches restrict the metasurface design and its minimum feature size. First, electrical switching between A- and C-PCMs requires implementation of electrodes, which in turn sets bounds for the metasurface design.[33] Second, local crystallization with optical pulses is limited by the transmission of the applied materials and, more importantly by diffraction of the laser light used for writing. Therefore, generating structural patterns in PCMs below ~λ/2 with λ being the laser wavelength is extremely challenging. If a reliable approach were available, deep sub-diffraction-limited switching of phase-change thin-films would enable new possibilities for the design and function of PCM-based metasurfaces.

Over the last decades progress has been made by reducing λ and increasing the numerical aperture of the lens(es) used, which increased the recording density from compact discs (CDs) to digital versatile discs (DVDs) and Blu-ray discs.[34] Furthermore, substantial efforts have been made to achieve control over the size of the switched areas in the metasurfaces,[23] but only C-PCM lateral sizes ≥ 500 nm have so far been realized.[22,23,30] However, unconventional switching by using carbon-nanotube electrodes,[35] conductive atomic force microscopy,[17] or heated scanning probes[36-38] have shown the potential to push the size limits far below the diffraction limit.

Here, we propose using thermal scanning-probe-induced local switching of a PCM layer[38] to realize near-infrared metasurfaces with feature sizes far below the diffraction limit. Such a fabrication technique would allow creation of re-programmable phase-change metasurfaces with ultra-high resolution. The metasurface design studied here consists of an array of C-PCM cylindrical disks written in the A-PCM thin film. This structure shows tunable perfect absorption in the NIR spectral range. Controlled tip-induced crystallization of the PCM layer can allow for broad-band tuning of the perfect absorption. We first discuss the optical properties of such an array assuming an idealized, strictly cylindrical geometry of C-PCM regions. Systematic analysis of the resonant modes contributing to the electromagnetic response of the metasurfaces is performed and complemented by direct numerical simulations. We further provide a detailed theoretical description of the tip-induced crystallization by employing multiphysics simulations. We demonstrate the great potential of the thermal scanning-probe technique in fabricating reconfigurable metasurfaces. Moreover, we derive guidelines for deeply subwavelength-sized photonic metasurfaces applicable to other materials providing a substantial refractive index change $\Delta n$. Our concept allows for tunable perfect absorption, and can be applied to sensing, thermal imaging, and spatial frequency filtering.[04,39]

## RESULTS

Thermal scanning-probe lithography enables high-resolution patterning and can provide optical surfaces with novel functionalities.[40,41] Since thermal scanning-probe lithography is typically done by patterning polymer surfaces, the direct applicability of this technique to switching PCMs is non-trivial. However, reproducible switching of micrometer-sized areas, (curved) lines, and small dots has been shown.[36-38] Here, we assume systematic application of the heated scanning probe, resulting in an array of C-PCM disks in a homogeneous A-PCM layer. Metasurfaces based on disk arrays have been demonstrated for the MIR,[42,43] NIR,[44-47] and terahertz[48] spectral range. These dielectric[42,43,47] and plasmonic resonators[44] showed multipolar resonances,[43] chiral sensing capabilities,[47] phase tuning,[42] and perfect absorption.[44,46] Due to their straight-forward design and the versatility of the device, we chose to study similar metasurfaces, hence our arrays are composed of C-PCM disks in an A-PCM film.

**Figure 1(a)** shows the design of our phase-change metasurface: a thin film stack consisting of a silicon (Si) substrate and silicon dioxide (SiO$_2$) as sandwich layers for the encapsulated PCM film. We selected germanium telluride (GeTe) as the PCM due to its optical and thermal properties. Further discussion and a comparison to Ge$_2$Sb$_2$Te$_5$, which is the PCM most often applied in photonics and storage media, is given in the SI. Si as well as SiO$_2$ were selected for thermal reasons with the former being an effective heat sink and the latter offering a thermal conductivity $\kappa$ almost identical to that of GeTe (*cf.* SI). Thus, the layer stack guarantees the thermal transport necessary for the amorphous-to-crystalline phase transition (and *vice versa*).[27] To protect the PCM from evaporation and oxidation, the top SiO$_2$ is used as a thin capping layer.[34,49] For the excitation of plasmonic modes, we added a thin gold (Au) film below the PCM.

The layer stack shown in Figure 1(a) can be understood as a Dallenbach perfect absorber consisting of homogeneous lossy dielectric layers (SiO$_2$|PCM|SiO$_2$) on top of a metallic ground plane (Au).[50] By carefully choosing the thickness of each layer, the reflected light can be suppressed via impedance matching. The resulting perfect absorption is then based on the concept of an asymmetric Fabry-Pérot (FP) cavity.

The thickness of the GeTe and gold layer was varied first to achieve perfect absorption with zero reflectance ($R = 0$), and second to shift the resonance position $\lambda_{res}$ to the short-$\lambda$ end of the telecommunication band. The shift $\Delta\lambda_{res}$ is caused by the strong refractive index increase and the related redshift expected upon crystallization of the GeTe film. Third, we kept the films as thin as possible. By applying these three principles to the GeTe and gold film, we optimized our metasurface design. **Figure 1(b)** shows that the use of a thinner A-GeTe layer has two effects. On one hand, the reflectance minimum $\lambda_{res,A}$ is shifted to shorter wavelengths, and on the other hand, the minimum reflectance is lowered from $R = 0.11$ to $R = 0$. A similar trend can be found for C-GeTe (*cf.* SI). Hence, we fixed $h_{A-GeTe} = 40$ nm and varied the thickness of the gold film $h_{Au}$ [**Figure 1(c)**]. Since the reflectance remained zero for all studied thicknesses $h_{Au}$ (*cf.* SI), we concentrated on the resonance position $\lambda_{res,A}$. The increase of $h_{Au}$ leads to decreasing $\lambda_{res,A}$, while no spectral shift is seen for $h_{Au} > 50$ nm. Thus, we chose $h_{Au} = 50$ nm for all calculations below.

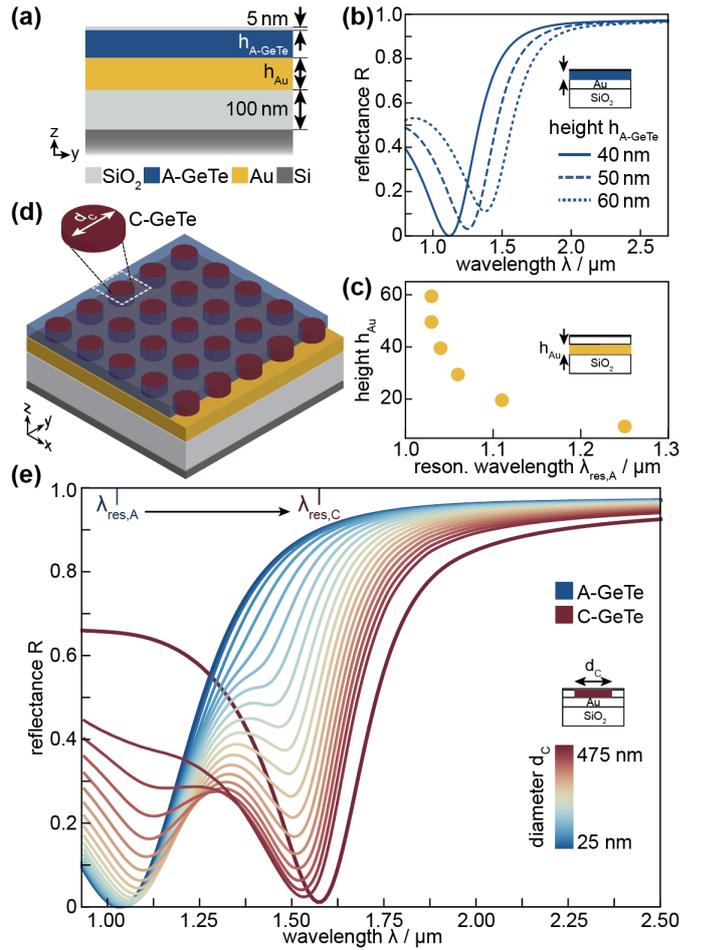

**Figure 1. Tunable phase-change metasurface. (a)** Planar thin-film stack including a silicon substrate, silicon dioxide thermal buffer, GeTe film, gold film, and silicon dioxide capping layer (bottom to top, color coded). The thin-film stack was optimized for perfect absorption and application at telecom wavelengths by varying the height of, **(b)**, the A-GeTe (blue), and, **(c)**, the gold layers. Since the resonance minimum $\lambda_{res,A}$ for a fully A-GeTe film (*i.e.* $d_C = 0$) is shifted to shorter wavelengths by decreasing $h_{A-GeTe}$ and increasing $h_{Au}$, we chose the final thicknesses $h_{GeTe} = 40$ nm and $h_{Au} = 50$ nm. **(d)** Phase-change metasurface with crystalline disks (red, diameter $d_C$) arranged in a square lattice with 500 nm periodicity (unit cell marked by dashed white square). We neglect silicon dioxide and partially remove the opaque A-GeTe for clarity. **(e)** Calculated spectra of the reflectance amplitude for different $d_C$ ranging from 25 to 475 nm in steps of 25 nm show tuning of the resonance features. For comparison, reflectance spectra for the fully A- and C-GeTe layer (thick dark blue and red lines, respectively) are shown. The increase in $d_C$ represents the total increase of the crystalline fraction in the GeTe film, thus shifting the spectra between $\lambda_{res,A}$ and $\lambda_{res,C}$ (top x-axis ticks). This is reflected qualitatively in the color code ranging from blue tones (rich in A-GeTe) to red tones (rich in C-GeTe).

**Figure 1(d)** shows the design of our metasurface. Local crystallization leads to C-disks placed in the center of each unit cell with 500 nm side length. The disks are defined by their diameter $d_C$. Here, we assume that the C-PCM disks penetrate over the entire GeTe layer thickness and have idealized cylindrical shape. We calculated the spectral reflectance $R$ of the metasurface for light in the NIR, as shown in **Figure 1(e)**. First, we determined $R$ for a fully amorphous and crystalline thin-film stack, i.e. no local phase change was considered. For this purpose, we applied the generalized transfer-matrix method;[51] the spectra are given in dark blue and dark red for A- and C-GeTe, respectively. Upon crystallization of the entire GeTe film, the resonance related to the perfect absorption shifts by $\Delta\lambda_{res} = 540$ nm from $\lambda_{res,A} = 1030$ nm to $\lambda_{res,C} = 1570$ nm. Further, we calculated spectral reflectance for different C-disk sizes $d_C$ by using full-wave 3D simulations (CST Microwave Studio®, cf. SI). The color of these spectra indicates if the layer stack is dominated by the A- or C-type multilayer reflectance, i.e. spectra for small or large $d_C$ are given in blue or red tones, respectively. We considered metasurfaces with $d_C$ ranging from 25 to 475 nm in steps of 25 nm. While all spectra reveal the dominance of the perfect absorption related to A- and C-GeTe multilayer, the introduction of the periodic disk array into the multilayer stack leads to one additional spectral feature that interacts with the dominant reflectance dip. For $d_C \geq 100$ nm, a local reflectance minimum ($\lambda = 1355$ nm for $d_C = 100$ nm) appears on the long-wavelength shoulder of the dip ($\lambda = 1041$ nm for $d_C = 100$ nm). With increasing $d_C$, this local minimum is shifted to longer wavelengths (e.g. $\lambda = 1481$ nm for $d_C = 250$ nm) and becomes more pronounced, i.e. $R$ decreases. Simultaneously, the main resonance dip shifts from $\lambda_{res,A}$ to longer wavelengths (e.g. $\lambda = 1070$ nm for $d_C = 250$ nm). Furthermore, the reflectance of this dip starts to increase for $d_C \geq 300$ nm. In the case of a metasurface with disks of $d_C = 375$ nm, both minima show $R \approx 0.12$ while still being spectrally separated by about 400 nm ($\lambda = 1106$ nm and $\lambda = 1511$ nm). For larger disks, the trend continues, and the long-wavelength reflectance dip becomes dominant. Thus, a local reflectance minimum ($\lambda = 1126$ nm for $d_C = 425$ nm) appears on the short-wavelength shoulder of the main reflectance dip ($\lambda = 1521$ nm for $d_C = 425$ nm). The change of the dominant absorber matches the ratio of the GeTe phases. For $d_C \leq 375$ nm, the majority of the GeTe in the metasurface is in its amorphous state while for $d_C = 400$ nm, already more than 50% of the GeTe are crystalline (cf. SI). The spectra for each diameter are shown individually in the SI.

The mentioned resonance features can be described as a combination of four different resonant modes, illustrated in **Figure 2**. The layer stack exhibits two vertical FP cavity modes — one related to the A-GeTe film, the other associated with the C-GeTe film [cf. **Figure 2(a)**]. By introducing a square array of crystalline disks, we simultaneously added an array of circular slot resonators for surface plasmon polaritons (SPPs) at the A-GeTe|Au interface [**Figure 2(b)**, left] and an array of circular disk resonators for SPPs at the C-GeTe|Au interface [Figure 2(b), right]. This results in a pair of horizontal SPP modes. In the metasurface spectra in **Figure 2(c)**, the perfect absorption features are visible. Furthermore, it can be seen that the disk and slot array SPP modes couple to the spectrally nearest perfect absorber mode. In contrast, the two vertical modes, Figure 2(a), cannot couple to each other due to spectral separation. The same holds for the horizontal SPP mode pair, Figure 2(b).

The spectral behavior of the SPP modes can be understood by employing a simple analytical model. The resonance condition for the SPP mode at the A-GeTe|Au interface is given by a standing wave condition at the circumference of the circular slot with diameter $d_C$:[53,54]

$$\beta_{SPP,A} = \frac{m + \varphi}{d_C} \quad (1)$$

with $m$ being an integer and $\varphi$ being the phase shift accounting for imperfect reflections at the A-/C-GeTe interfaces. The resonance condition for the SPP mode at the C-GeTe|Au interface is given by the Bessel-type standing-wave condition inside of the circular disk:[54,55]

$$\beta_{SPP,C} = \frac{x_j(J_k) + \varphi}{d_C} \quad (2)$$

with $x_j$ denoting the $j$-th zero-crossing root of the $k$-th Bessel function of the first kind, $J_k$. In equations (1) and (2), $\beta_{SPP,A/C}$ is a real wavevector[56] given by the dispersion relation of the SPPs bound to the corresponding GeTe|Au interface in the analyzed multilayer.[57,58] The considered dispersion relations have been calculated using generalized transfer-matrix method[51] and are shown in the SI.

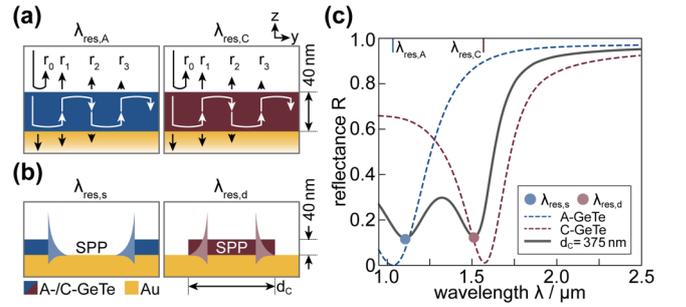

**Figure 2. Overview of resonant modes. (a)** The vertical modes are related to the thin-film interferences in GeTe schematically shown in the top with $r_0$, $r_1$, $r_2$, $r_3$ being partially reflected waves at the A-/C-GeTe film (left/right).[52] **(b)** The horizontal SPP modes can be ascribed to the slot in the A-GeTe film, left, and the C-GeTe disk, right. **(c)** The reflectance spectrum for a phase-change metasurface with disk diameter $d_C = 375$ nm (gray line) is compared to the spectra for fully A-/C-GeTe films (dashed blue and red line). The metasurface spectrum shows the influence of all four modes: the vertical modes, characterized by $\lambda_{res,A}$ and $\lambda_{res,C}$, due to the bare GeTe film, and the horizontal SPP modes with resonance wavelengths $\lambda_{res,s}$ and $\lambda_{res,d}$.

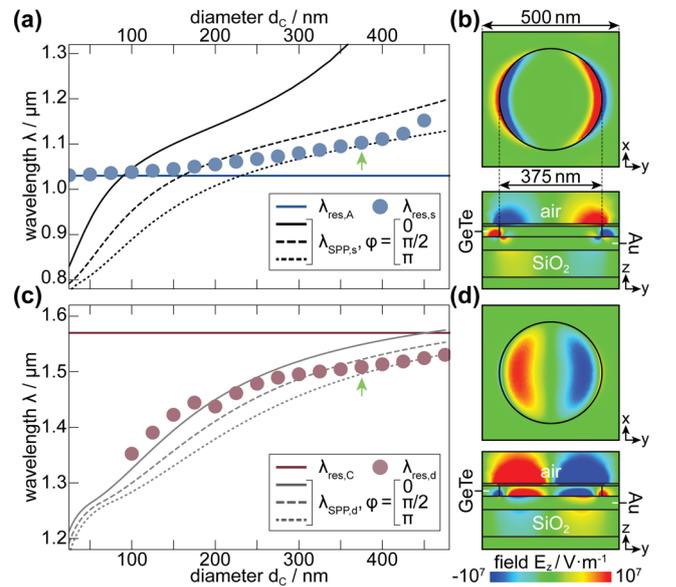

**Figure 3. Mode interplay. (a)** Short-wavelength resonance positions $\lambda_{res,s}$ extracted from calculated reflectance spectra in Figure 1(e) are given by blue dots. The perfect absorber resonance for fully A-GeTe film $\lambda_{res,A}$ is shown as a blue straight line. The analytical description based on SPP resonances $\lambda_{SPP,s}$ (black lines) describes $\lambda_{res,s}$ well for a phase shift $\varphi = \pi$ (dotted line). **(b)** The simulated field distribution of Re($E_z$) for $d_C = 375$ nm at resonance $\lambda_{res,s} = 1.10$ μm [green arrow in (a)] is confined to the disk edge. **(c)** Long-wavelength resonance positions $\lambda_{res,d}$ extracted from Figure 1(e) are given by red dots with $\lambda_{res,C}$ shown as red straight line. The analytical description based on SPP resonances $\lambda_{SPP,d}$ (gray lines) describes $\lambda_{res,d}$ well for different phase shifts $\varphi$. **(d)** The calculated field distribution Re($E_z$) for $d_C = 375$ nm at resonance $\lambda_{res,d} = 1.51$ μm [green arrow in (c)] is confined to the inside of the disk.

Solving resonance conditions (1) and (2) with the corresponding dispersion relation results in a set of complex frequencies ω of the SPP modes for a given slot/disk diameter with the spectral position given by:[58]

$$\omega_{SPP,s/d} = \sqrt{(\mathrm{Re}[\omega])^2 + (\mathrm{Im}[\omega])^2} \quad (3a)$$

$$\lambda_{SPP,s/d} = \frac{2\pi c}{\omega_{SPP,s/d}} \quad (3b)$$

In **Figure 3** we compare the analytical model (1-3) with numerical results. Please note that not all modes can be excited in the system by the used excitation (linearly polarized plane wave at normal incidence). The short-wavelength (long-wavelength) resonance positions $\lambda_{res,s}$ ($\lambda_{res,d}$) extracted from the calculated reflectance spectra in Figure 1(e) are given by blue (red) dots, and the solution of the analytical model $\lambda_{SPP,s}$ ($\lambda_{SPP,d}$) is given as black (grey) lines for slot (disk) SPP modes. First, we focus our discussion on the resonances $\lambda_{res,s}$ found at shorter wavelengths, shown in **Figure 3(a)**. Here, avoided crossing of the vertical and horizontal modes can be expected. However, due to the increasing absorption of GeTe for decreasing λ (*cf.* SI), only the upper branch can be observed in the numerical simulations. For small $d_C$, the perfect absorber wavelength $\lambda_{res,A}$ is dominant. The dominance is weakened with $\lambda_{res,s}$ departing from $\lambda_{res,A}$, which is the case for medium sized disks. Instead, $\lambda_{res,s}$ approximately follows the SPP slot mode. Here, the best fit is obtained for the second-order mode ($m=2$) and the phase shift φ = π. The simulated field distributions for a disk array with $d_C$ = 375 nm at resonance [**Figure 3(b)**] confirms our model assumption: the field is distributed at the edge of the circular slot cavity at the A-GeTe|Au.

The relation between $d_C$ and the resonance positions $\lambda_{res,d}$ found at longer wavelengths is shown in **Figure 3(c)**. As mentioned before, $\lambda_{res,d}$ appears for $d_C \geq 100$ nm and becomes more pronounced with increasing disk diameter. For $x_1(J_2)$, the analytical model with $\lambda_{SPP,d}$ matches the resonance positions $\lambda_{res,d}$ very well as can be seen in Figure 3(c). Nevertheless, the spectral separation between the Bessel-type SPP resonances $\lambda_{res,d}$ and the perfect absorption feature at $\lambda_{res,C}$ leads to coupling between both only for larger diameters $d_C$. Thus, a flattening of the $\lambda_{res,d}(d_C)$ trend relative to the $\lambda_{SPP,d}(d_C)$ behavior can be observed. The simulated field distributions in **Figure 3(d)** confirms the hypothesis of a Bessel-type SPP. This field plot reveals a $TEM_{01}$ mode in the C-GeTe|Au disk which has been observed for surface phonon polaritons in circular gold cavities as well as for hBN nanodisks, both featuring resonances in the MIR.[31,59] The given analytical descriptions of the resonances ascribed to the C-GeTe|Au disk and A-GeTe|Au slot array verifies the assumed modes present in the metasurface, as illustrated in Figure 2.

In contrast to other metasurface-based perfect absorbers in the literature, our design allows for the shift of the absorption band without any change of the layer stack, *i.e.* fabrication of further samples is not necessary.[60,61] However, the realization of such small crystallized regions in the amorphous matrix is challenging with commonly applied methods. Here, we propose to use a thermal scanning probe to induce local switching.[38] A few promising results showed the potential of using a heated tip for locally crystallizing the surface of amorphous $Ge_2Sb_2Te_5$[36] and GeTe films.[37] However, difficulties with controlling the depth (switching in *z*-direction) were reported and tip-induced re-amorphization has not yet been demonstrated.[36,37] More importantly, tunable optics based on such high-resolution switching have not been discussed.

To assess the potential of tip-induced local switching of PCM thin films we applied a self-consistent multiphysics description to model the crystallization kinetics under the influence of a heated scanning probe.[27] While we formerly used this description for laser-induced switching of $Ge_2Sb_2Te_5$,[27] we extended the model to tip-induced crystallization of GeTe (*cf.* SI). Since the metasurface design is based on a C-GeTe disk array, local crystallization of periodic dots is needed. These patterns can be generated by placing the tip, which is heated to *T* = 1223 K, in the center of the target position, as shown in **Figure 4(a)**. The size and volumetric extension of the crystallized zones can be selected based on the sample-probe contact duration (dwell time *t*). We conducted a dwell-time-dependent study of the crystallized diameter at the top and bottom of the GeTe film $d_{C,T}$ and $d_{C,B}$, respectively. The continuous time dependency of $d_{C,T}$ in **Figure 4(b)** reveals a strong increase of the top diameter during the first 50 ns of sample-probe contact. Afterwards, the curve flattens at about 30% of the array periodicity. In contrast, the bottom diameter $d_{C,B}$ remains close to zero for the first 350 ns of sample-probe contact, while it increases strongly after *t* = 50 μs [**Figure 4(c)**]. The snapshots in **Figure 4(d)** illustrate the *t*-dependent evolution of the GeTe crystallization. Not only does $d_{C,T}$ increase, but also the crystallized volume changes significantly and develops a lens-like shape. Crucial for optoelectronic applications is the fact, that the crystallization front reaches the substrate on the microsecond scale, enabling interaction with the (photonic) structures or plasmonic material below.

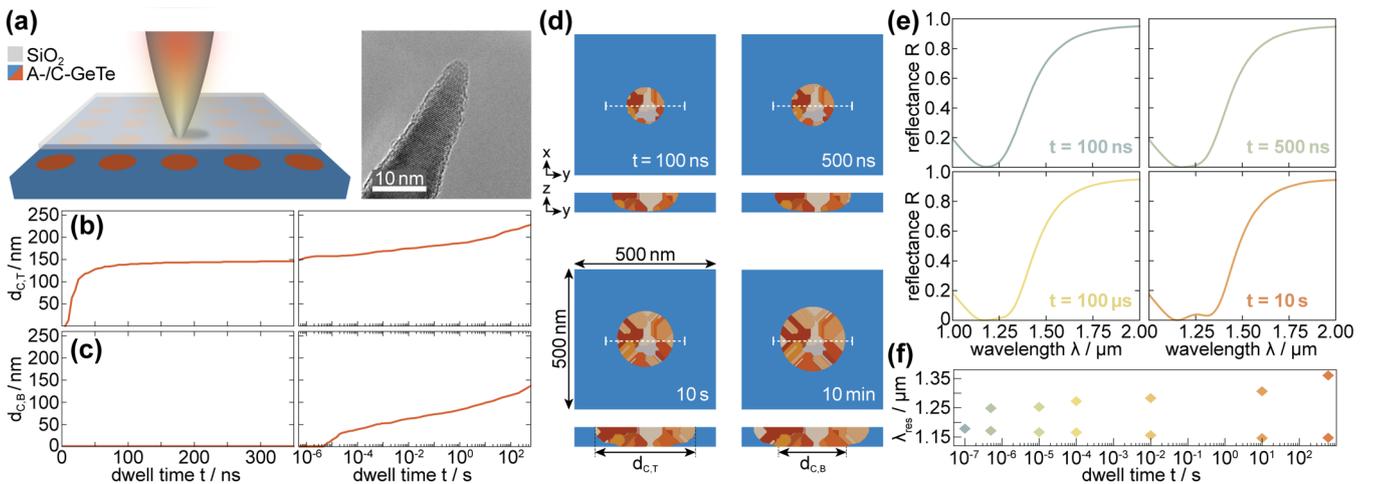

**Figure 4. Modeling of tip-induced local crystallization of the metasurface. (a)** Local crystallization with a heated tip leads to C-GeTe disks in a 40 nm thick A-GeTe film (capped with silicon dioxide). The high-resolution transmission electron microscopy image shows a typical tip before patterning (courtesy of Tevis D. B. Jacobs, University of Pittsburgh). The diameter of the crystallized spot at the top and bottom of the GeTe film $d_{C,T}$ and $d_{C,B}$ in **(b)** and **(c)**, respectively, depends on the dwell time. While $d_{C,T}$ increases strongly for the first 50 ns of tip–sample contact, this behavior starts to flatten for longer *t*. At the bottom of the GeTe, the diameter of the crystallized spot $d_{C,B}$ increases very slowly. **(d)** Snapshots of the crystallization in the GeTe film induced by the heated tip after 100 ns, 500 ns, 10 s, and 10 min dwell time; the crystallized volume (composed of differently oriented grains shown in orange shades) increases steadily. The dashed white line across the crystalline zone in each top view marks the region given in the cross sections below. **(e)** Calculated reflectance spectra for the dwell times evaluated in (d) with **(f)** showing the resonance positions $\lambda_{res}$ for different dwell times (color code as for the spectra). All related reflectance spectra are shown in the SI.

For local crystallization induced by the heated tip in the form of the periodic array (500 nm periodicity), we calculated the reflectance spectra as shown in **Figure 4(e)**. Similar to the reflectance spectra calculated for an idealized structure in Figure 1(e), one broad reflectance dip with $R = 0$ occurs for small crystalline volumes ($t = 100$ ns). After $t = 500$ ns, crystallization progresses. Thus, we see two dips and a resonance shift $\Delta\lambda_{res} = 71$ nm [**Figure 4(f)**]. This trend continues for larger crystalline volumes caused by longer dwell times $t$. However, the calculated spectra shown in Figure 4(e) differ from the spectra given in Figure 1(e), especially for shorter $t$. This can be ascribed to the discrepancy between idealized and actual shape of the crystallized regions as well as to the introduction of a thermal buffer layer between GeTe and gold for the multiphysics simulation (*cf.* SI). This layer was necessary to achieve a greater crystallization depth in a shorter time frame. A further decrease of the dwell time could be realized by optimization of the thermal design of the layer stack in combination with a variable tip-layer stack contact area, corresponding to the desired diameter of the crystalline disks.

Overall, the multiphysics simulation demonstrates that the tip-induced local switching could be used to create phase-change metasurfaces with an ultra-high resolution on a microsecond timescale. The erasure of the created crystalline features can be realized by single femtosecond laser pulses which allow for re-amorphization of PCM areas with a diameter larger than 200 μm.[21] Recently, combining thermal scanning probe lithography with a laser writer has been used to demonstrate a hybrid lithography approach that simultaneously can achieve high-resolution and high-throughput patterning.[62] If the appropriate laser were to be combined with the thermal scanning probe, it could be employed for both high-throughput switching of large-area PCM patterns, as well as for re-amorphization, enhancing the capabilities of our proposed approach.

## CONCLUSION

PCMs have been shown to be a great platform for tunable optics, ranging from hybrid plasmonic to all-dielectric metasurfaces and from on-chip photonic memories to displays. For most of these tunable optical elements, maximizing the spatial confinement is beneficial not only to reduce device dimension and therefore, increase information density, but also to reduce energy consumption. The smaller the PCM volume that has to be switched, the less power is needed to induce the phase transition.

Here, we explored a metasurface design based on ultra-small tunable features in GeTe allowing for a tunable perfect absorber. We proposed using thermal scanning-probe-induced local switching to realize metasurfaces with feature sizes far below the diffraction limit. We chose sub-diffraction limited crystalline patterns in amorphous GeTe films on a thin gold film with a design optimized for the telecommunication bands situated in the NIR spectral range. Our sample architecture utilizes three-dimensionally confined structures to strongly localize the electric field. In contrast to other PCM-based metasurfaces, the spectral tuning of the perfect absorption of our design is not based on multiple samples each with different resonator sizes[60,61] and thus, extremely versatile.

Due to the diffraction limit, laser-induced switching of PCMs inhibits the miniaturization of (existing) designs. Thus, alternative ways for local switching need to be exploited. Here, we presented multiphysics simulations of tip-induced switching of the PCM-based metasurface. This allowed us to predict local crystallization in the GeTe film with feature sizes down to tens of nanometers. Additionally, our calculated results explain the limited switching depths reported in experimental studies on tip-induced crystallization,[36,37] especially considering the lower thermal conductivity of crystalline GST compared to GeTe.[36] Thus, our insights give instructions for tip-induced crystallization of PCMs and therefore, pave the way towards ultra-small and high-density tunable photonics. Furthermore, the concepts can be transferred to optical storage media, thermal imaging, spatial-frequency filtering, and biosensing.[04,36,39,63]

## AUTHOR INFORMATION

### Corresponding Authors

**Ann-Katrin U. Michel** – *Optical Materials Engineering Laboratory, Department of Mechanical and Process Engineering, ETH Zurich, 8092 Zurich, Switzerland*
orcid.org/0000-0002-6130-3991; Email: micheann@ethz.ch
**Dmitry N. Chigrin** – *DWI – Leibniz Institute for Interactive Materials, 52076 Aachen, Germany; I. Physikalisches Institut (1A), RWTH Aachen University, 52056 Aachen, Germany*
orcid.org/0000-0002-8197-707X; Email: chigrin@ dwi.rwth-aachen.de

### Authors

**Sebastian Meyer** – *I. Physikalisches Institut (1A), RWTH Aachen University, 52056 Aachen, Germany*
orcid.org/0000-0003-4208-0988
**Nicolas Essing** – *I. Physikalisches Institut (1A), RWTH Aachen University, 52056 Aachen, Germany*
**Nolan Lassaline** – *Optical Materials Engineering Laboratory, Department of Mechanical and Process Engineering, ETH Zurich, 8092 Zurich, Switzerland*
orcid.org/0000-0002-5854-3900
**Carin R. Lightner** – *Optical Materials Engineering Laboratory, Department of Mechanical and Process Engineering, ETH Zurich, 8092 Zurich, Switzerland*
**Samuel Bisig** – *Heidelberg Instruments Nano/SwissLitho, Technoparkstrasse 1, 8005 Zurich, Switzerland*
**David J. Norris** – *Optical Materials Engineering Laboratory, Department of Mechanical and Process Engineering, ETH Zurich, 8092 Zurich, Switzerland*
orcid.org/0000-0002-3765-0678

### Conflicts of interest

There are no conflicts to declare.

## ACKNOWLEDGMENT

The authors thank Abu Sebastian, Iason Giannopoulos, and Vara Prasad Jonnalagadda from IBM Zurich, as well as Raphael Brechbühler from ETH Zurich for helpful discussions. A.U.M. acknowledges funding from the ETH Zurich Postdoctoral Fellowship Program and the Marie Curie Actions for People COFUND Program (Grant 17-1 FEL-51). This project was partially funded by the European Research Council under the European Union's Seventh Framework Program (FP/2007-2013)/ERC Grant Agreement Number 339905 (QuaDoPS Advanced Grant). D.N.C. acknowledges a partial support by the DFG through the Heisenberg Fellowship (CH 407/7-2).